\documentclass[twocolumn,superscriptaddress,longbibliography]{revtex4-2}
\setcitestyle{numbers,square}

\usepackage{braket}
\usepackage{graphicx}
\usepackage{dcolumn}
\usepackage{bm}
\usepackage[normalem]{ulem} 
\usepackage{color}
\usepackage{siunitx}
\usepackage{braket}
\usepackage[version=4]{mhchem}
\DeclareSIUnit\elementarycharge{\text{\ensuremath{e}}}
\DeclareSIUnit\angstrom{\text {Å}}
\DeclareSIUnit\uc{\text {unit-cell}}
\usepackage{hyperref}
\hypersetup{
    colorlinks,%
    citecolor=blue,%
    linkcolor=blue,%
    urlcolor=blue
}

\newcommand{\orcid}[1]{\href{https://orcid.org/#1}{\includegraphics[width=8pt]{orcid.pdf}}}

\begin{document}

\title{Theory of spin and orbital charge conversion at the surface states of Bi$_{1-x}$Sb$_x$ topological insulator}

\author{Armando Pezo}
\affiliation{Laboratoire Albert Fert, CNRS, Thales, Université Paris-Saclay, 91767, Palaiseau, France}

\author{Jean-Marie George} 
\affiliation{Laboratoire Albert Fert, CNRS, Thales, Université Paris-Saclay, 91767, Palaiseau, France}

\author{Henri Jaffrès} 
\affiliation{Laboratoire Albert Fert, CNRS, Thales, Université Paris-Saclay, 91767, Palaiseau, France}

\date{\today}

\begin{abstract}
Topological insulators are quantum materials involving Time-reversal protected surface states (TSS) making them appealing candidates for the design of next generation of highly efficient spintronic devices. The very recent observation of large transient spin-charge conversion (SCC) and subsequent powerful THz emission from Co$|$Bi$_{1-x}$Sb$_x$ bilayers clearly demonstrates such potentiality and feasibility for the near future.  Amongst the exotic properties appearing in and at the surface of such quantum materials, spin-momentum locking (SML) and Rashba-Edelstein effects remain as key ingredients to effectively convert the spin degree of freedom into a charge or a voltage signal. In this work, we extend our analyses to the quantification of orbital momentum-locking and related orbital charge conversion effects in Bi$_{.85}$Sb$_{.15}$ via orbital Rashba-Edelstein effects. In that sense, we will provide some clear theoretical and numerical insights implemented by multiorbital and slab tight-binding methods (TB) to clarify our recent experimental results obtained by THz-TDS spectroscopy.
\end{abstract}

\maketitle

\section{Introduction} 
Theoretical proposals made almost two decades ago for exotic materials displaying an insulating bulk with metallic surfaces states~\cite{kane_mele_qshe_graphene,Bernevig1757} led quickly to their experimental observation by measuring the spin Hall conductance in HgTe|CdTe quantum wells~\cite{Konig} and more recently in two-dimensional materials like bismuthene~\cite{Reis2017}. The strong spin-orbit coupling (SOC) in Bi-based materials makes them ideal candidates for spintronics and  valleytronics applications~\cite{Ji2016, Lu2017, Zhang2015, Liu2017, Wang2017_1} owing to the so-called band inversion mechanism responsible for the emergence of their topological properties. Among them, the Bulk-Boundary correspondence~\cite{essin_bulk_boundary}, relates their topological classification to the existence of spin-polarized surface states (TSS)~\cite{z2_KaneMele.95.146802} manifesting a strong spin momentum locking (SML) preventing back-scattering as long as disorder does not break heavily the time reversal symmetry (TRS)~\cite{PhysRevMaterials.5.014204,Pezo2023}. In a couple of recent papers, it was successfully demonstrated by THz spectroscopy in the time domain (THz-TDS) that Ag$|$Bi Rashba~\cite{Jungfleisch2018}, Bi$_2$Se$_3$~\cite{Chia2018} and SnBiTe TI~\cite{Rongione2022} as well as Bi$_{1-x}$Sb$_x$ alloy family~\cite{Sharma2021,Park2022,Rongione2023,rho2023} enables robust spin-charge conversion (SCC) from 3\textit{d} ferromagnetic (FM) injectors, proven to be as efficient as usual Pt or W heavy metals owing to the particular six-fold symmetry SML displayed by their TSS. Moreover, recent ab-initio calculations emphasize on the optical generation of orbital currents in BiAg$_2$ surface~\cite{Mokrousov2024}.

The rise of the orbital angular momentum (OAM) from electronic quasiparticles as a new degree of freedom has recently attracted much attention~\cite{Go2017,Go_orbital_texture,pezo_ohe}. This is partly explained by the ability to generate a prominant orbital angular momentum flow without the restricting requirement of a large spin-orbit strength; and possibly largely exceeding the spin flow generated by Pt or W~\cite{PhysRevB.77.165117}. It was postulated that either the orbital Hall effect (OHE) or the orbital Rashba-Edelstein effect (OREE), as well as the reciprocal inverse orbital Rashba-Edelstein effect (iOREE), may arise from the orbital texture in the bulk or from the orbital momentum locking (OML) at interfaces even involving centrosymmetric materials~\cite{PhysRevLett.121.086602}. Conversely, the occurrence of a spin momentum locking on the Fermi surface under the action of the spin-orbit coupling suggests a chiral OML texture, as previously demonstrated by angular-resolved photo emission (ARPES) on Bi$_2$Se$_3$~\cite{Taniguchi2012}, and possibly resulting in an orbital to charge conversion phenomenon. These ingredients bring new avenues as it broadens the set of materials and also poses new challenges owing to the intrinsic entanglement with the spin degree of freedom. In this regard, one way to reduce the role of the spin-orbit coupling is to use light materials as LaAlO$_3$$|$SrTiO$_3$~\cite{ElHamdi2023}, KTaO$_3$$|$Al~\cite{Varotto2022}, CuO$_x$~\cite{Krishnia2024} oxides or Co$|$Al~\cite{krishnia2023} or Ni$|$Cu~\cite{Xu2024} metallic interfaces.

\vspace{0.1in} 

If spin and orbital degrees of freedom are both to be considered in the inverse Rashba-Edelstein process arising onto a 2-dimensional (2D) converter material, the production of the charge current density at the surface is expected to scale as: 

\begin{equation}  
v_{k_\parallel,n}|\braket{\phi^{neq}_{k_\parallel,l}|\psi^{int}_{k_\parallel,n}}|^2
\label{generalequation}
\end{equation}
for specific 2D $k_\parallel$ electronic wavevector in the Brillouin zone. This describes the different quantum transition probabilities $|\braket{\phi^{neq}_{k_\parallel,l}|\psi^{int}_{k_\parallel,n}}|^2$ between \textit{non-equilibrium } occupied electronic states in a certain orbital-symmetry band ($l$) $\phi^{neq}_{k_\parallel,l}$ possibly optically excited from a ferromagnetic reservoir, and launched onto a material surface or interface (\textit{int}) with wavefunctions $\psi^{int}_{k_\parallel,n}$ in an orbital-symmetry band ($n$). $v_{k_\parallel,n}$ represents the velocity of the band $(n)$ at the $k_\parallel$ point. Note that such quantum transition probability given above can be also recovered in the Hamiltonian transfer in the limit of a certain contact potential~\cite{Schwab_2011}. In that sense, $\ket{\phi^{neq}}$ represents the wavefunction of the excited carrier very close to the surface (or interface) after having crossed the injector material. By this way, the unbalanced probability for the excited electrons to relax onto the respective $\pm k_\parallel$ wavevector in the band $(n)$ is able to produce the observed current charge along the surface plane, the resulting current being closely dependent of the inner symmetry band properties. Consequently, the whole degrees of freedom involving now both the spin and orbital quantities (and not only spin) may be treated at equal footing in the conversion processes once the exact electronic wavefunctions are known. We then has to consider a possible generation of non-zero orbital moment and subsequent conservation of such quantities over typically short nanometric distance, typically nanometer metallic layer thickness, as shown recently through the use of ultrashort optical laser pulse exciting either Ni~\cite{Xu2024} or CoPt~\cite{Xu2024b} ferromagnetic materials. In that purpose, note that Eq.~[\ref{generalequation}] may be also written $\left(f_{n,l}\right) \left[v_{k_\parallel,n}\bra{\psi^{int}_{k_\parallel,n}}\hat{\Pi}^{neq}_{k_\parallel}\ket{\psi^{int}_{k_\parallel,n}}\right]$ with $\hat{\Pi}^{neq}_{k_\parallel}=\varrho_{k_\parallel}^{neq}=\Sigma_{(l)} \ket{\phi^{neq}_{k_\parallel,l}}\bra{\phi^{neq}_{k_\parallel,l}}$ the \textit{density matrix} of the excited state, projecting thus onto the \textit{interfacial} material states, formulation that we will extensively use in the next. Here, $f_{n,l}$, the Fermi function describing the occupied states $(n,l)$ lying on the same energy shell. 

\vspace{0.1in}

In that prospect, the aim of this theoretical work is twofold. We first target to estimate and possibly disantangle respective spin (section V) and orbital charge interconversion (section VI) efficiencies in Bi$_{1-x}$Sb$_x$ alloys TI known as a very efficient converter. In particular, we explore quantitatively the different mechanisms which may come into play among spin and orbital Hall effects arising from a specific Berry curvature or Rashba-Edelstein contributions, marking a difference between bulk (section III) and general interfacial REE phenomena(section IV). We compare hereafter our experiments to the results obtained from a tight-binding (TB) Hamiltonian we have developed for the description of ultrathin Bi$_{1-x}$Sb$_x$ layers for $x=0.15$ (details are given in section II). Furthermore, our calculations for bulk Bi using this TB approach were compared with those directly originated from Density Functional Theory (DFT) as portrayed in Fig.~\textcolor{blue}{[S1]} of the supplementary information \textcolor{blue}{SI-IIA} (section III); with however the advantage of the TB method to treat more easily the case of BiSb alloys than DFT may achieve. By this way, the description of the electronic properties is demonstrated to be well suited for such TI and this gives a fair evaluation of the surface state spin texture in close agreement with the one derived from more refined DFT methods (section III). In the end, we will estimate the impact of the overall exchange potential terms $\Delta_{exc}$ acting on the surface of BiSb TI and acting onto the respective spin and orbital charge conversion arising from the contact with the ferromagnetic contact (sections V and VI).

\section{Model TB Hamiltonian} 

Bi and Sb are both group five semimetals with a direct gap throughout the entire Brillouin zone but a negative indirect gap from band overlap. Indeed, at the L point in Bi
the valence band (VB) maximum consists of odd parity antisymmetric orbitals and the conduction band (CB) minimum consists of even parity symmetric orbitals with
Z$_2$ = (0; 000). In pure Sb the orbitals are inverted leading to a nontrivial bulk invariant Z$_2$ =(1; 111). By alloying Sb into Bi, the small gap at the L point inverts and opens leading to a topologically nontrivial bulk insulator~\cite{Hsieh2008,Hsieh2009,teo2008} leading to topological surface states (TSS) having their spin locked perpendicular to momentum.

\vspace{0.1 in}

We first describe our TB model Hamiltonian to reproduce the electronic band structure of Bi$_{1-x}$Sb$_{x}$ bare multilayers. To these ends, we used a modified $\{s,p_x,p_y,p_z\}\otimes \{\uparrow,\downarrow\}$ \textit{sp}$^3$ tight-binding (TB) multiband Hamiltonian~\cite{liu_tb_BiSb} for Bi~\cite{Hofmann2006} developed for thin films and capturing our ARPES data~\cite{Baringthon2022,Rongione2023} (see the supplementary information \textcolor{blue}{SI-I}). This approach was used to demonstrate how the non-trivial topological phase arises for a given Sb content $x$ in a certain window~\cite{Lenoir2001,teo2008}. The electronic properties of the BiSb alloys are derived from the modified Virtual Crystal Approximation (VCA)~\cite{PhysRevB.80.085307}. Likewise, our TB treatment as detailed hereafter well exhibit the expected band inversion for the bulk BiSb as well as  topological properties of Bismuthene (1BL Bi) unlike Sb (refer to the supplementary information \textcolor{blue}{SI-II)}. However, adequate surface potential terms, including possible contact exchange terms with the ferromagnetic layer, have to be possibly added to recover the proper electronic properties. The impact of an exchange contact will be more particularly discussed in the end of the paper.

\vspace{0.1in}

The rhombohedral A7 structure of Bi, Sb and BiSb alloys is described by two atoms per unit cell, forming then a bilayer (BL) of thickness of about 0.4~nm. Bi$_{1-x}$Sb$_x$ slabs are obtained by stacking the BL along the (1\,1\,1) direction ($z$ axis) with two different plane-to-plane distances. We constructed our Hamiltonian on the basis of the work of Ref.~\cite{saito2016} using the generalization of the \textit{sp$^3$} TB-model Hamiltonian proposed for pure bulk Bi and Sb crystals~\cite{liu_tb_BiSb}, and adapted to Bi$_{1-x}$Sb$_x$ alloys~ in Ref.~\cite{teo2008}. The interactions Hamiltonian is complemented by the introduction of additional surface potential terms when dealing with thin layers (treatment in slabs)~\cite{petersen2000,ast2012,saito2016}. In particular, the hopping parameters for the BiSb alloys are obtained by using the virtual crystal approximation (VCA) according to:

\begin{equation}
V_\text{C}^\text{BiSb}=x~V_\text{C}^\text{Sb}+(1-x^2)~V_\text{C}^\text{Bi}
\end{equation}
with $x$ is the Sb content and $V_\text{C}^\text{Sb}$ and $V_\text{C}^\text{Bi}$ are the respective hopping parameters of Sb and Bi taken from Ref.~\cite{liu_tb_BiSb}. 

\begin{figure}[ht!]
\includegraphics[width=\linewidth]{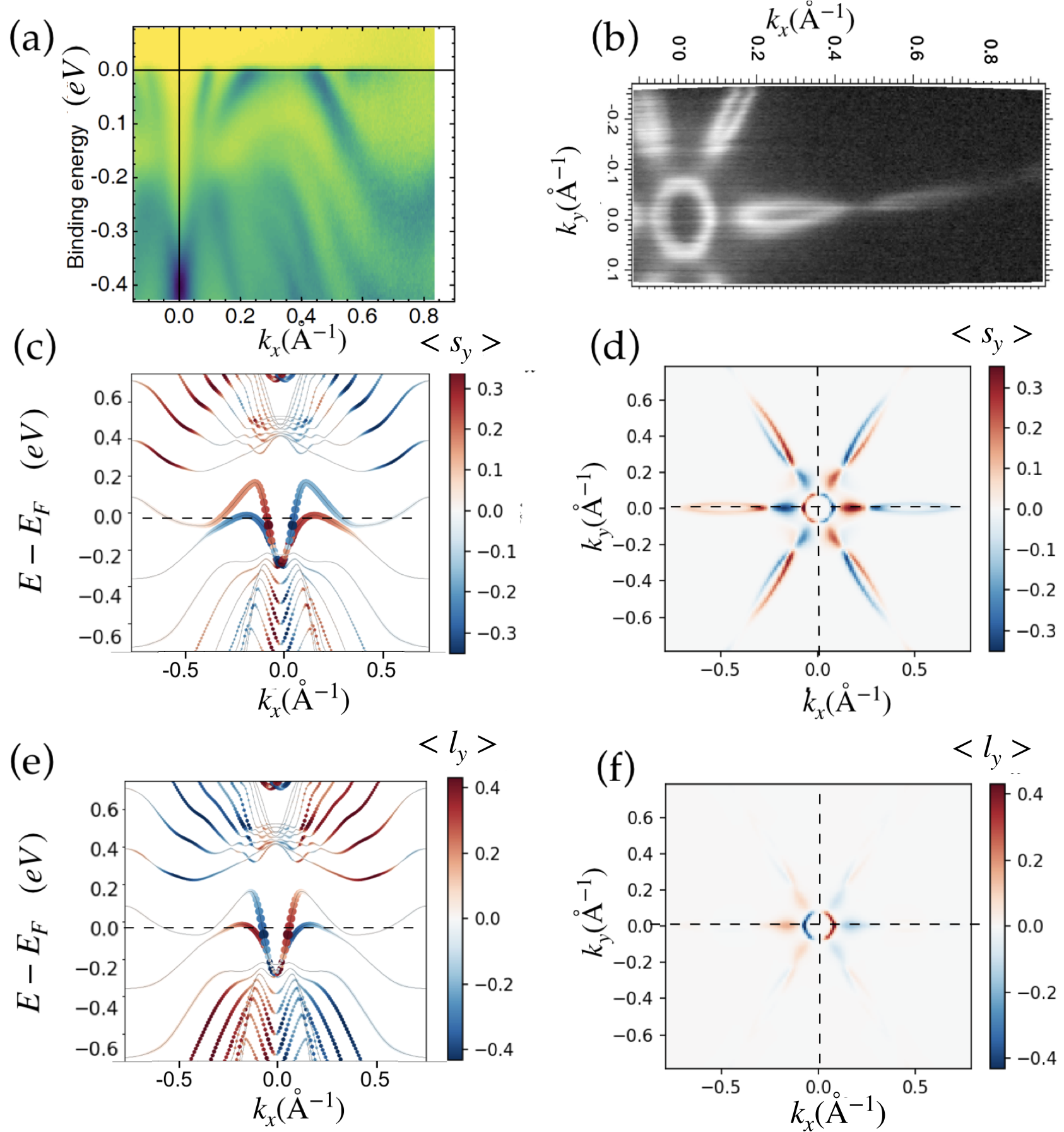}
\caption{\textbf{ARPES data and TB modelling of 12 BLs Bi$_{.85}$Sb$_{.15}$ films.} (a) ARPES data (from Ref.~\cite{Rongione2023}) acquired on a 5~nm (12 BLs) Bi$_{.85}$Sb$_{.15}$ film displaying electronic energy dispersion along the $\hat{x}=\overline{M}-\overline{\Gamma}-\overline{M}^{'}$ line in the ($E=-0.4,0$)~eV energy window ($E=0$ corresponds to the Fermi energy). (b) ARPES data of the 2D Fermi cut of the DOS. (c) and (e) TB calculation of the respective spin ($\sigma_y$) and orbital ($L_y$) DOS projection onto the first ($1^{st}$) BiSb BL along the same $\hat{x}=\overline{M}^{'}-\overline{\Gamma}-\overline{M}$ high symmetry line. The bottom BL shows identical energy dispersion with opposite angular momentum chirality. (d-f) TB calculation of the Fermi surface 2D cuts of the $\hat{y}$ component of respective spin and orbital resolved DOS projected onto the $1^{st}$ BL.}\label{fig:2d-plots}
\end{figure}

The different hopping terms among the atomic orbitals are decomposed into inter- and intra-BL hopping terms. The inter-BL off-diagonal hopping term between atoms (plane) 1 and atoms (plane) 2 consists of the nearest-neighbor coupling in the bulk BiSb Hamiltonian, whereas the intra-BL hopping term consists of two parts which represents respectively the third and second nearest neighbor contributions. We considered the overall TB Hamiltonian according to:

\begin{equation}
\hat{\mathcal{H}}=\hat{\mathcal{H}}_0+\hat{\mathcal{H}}_\text{SO}+\hat{\mathcal{H}}_\gamma + \hat{\mathcal{H}}_{exc}
\end{equation}
where $\hat{\mathcal{H}}_0=\Sigma_{i\mu}^{j\nu} \ket{i\mu} V_{i\mu}^{j\nu}\bra{j \nu}$ represents the hopping Hamiltonian ($i$, $j$ are the atomic positions, $\mu$, $\nu$ are the orbitals), $\hat{\mathcal{H}}_\text{SO}=\frac{\hbar}{4m^2c^2}\left(\overrightarrow{\nabla} V(r) \times \hat{p}\right) \cdot \hat{\sigma}$ the \textit{bulk} atomic spherically symmetric spin-orbit term. The latter writes simply $\hat{\mathcal{H}}_\text{SO}=\lambda_{SO}\hat{L}.\hat{s}$ ($\hat{L}$ and $\hat{s}$ are the respective vectorial orbital moment and spin operators and $\lambda_{SO}$ the spin-orbit strength). Moreover, $\hat{\mathcal{H}}_\gamma + \hat{\mathcal{H}}_{exc}=\hat{\mathcal{V}}_{int}(z)$ are additional surface potential terms which can be viewed as the necessary potential $\hat{\mathcal{V}}_{int}$ to add on each multilayer surfaces and imposed by the presence of a local electric field $\mathcal{E}_z$ associated to local dipole moments. 

\vspace{0.1in}

With more details, $\hat{\mathcal{H}}_{exc}= - \Delta_{exc} \hat{m}\cdot \Sigma_{i\mu\nu}^{s s^\prime} \ket{i\mu s}\left[\hat{\sigma}\right]_{ss^\prime}\bra{i\nu s^\prime}$ a possible on-site contact exchange term acting on the first TI bilayer (BL) at the direct interface with the FM contact (Co) of unit magnetization direction $\hat{m}$. We note $\hat{\sigma}_{x,y,z}=2\hat{s}_{x,y,z}$ the corresponding Pauli matrix and $\hat{s}$ the vectorial spin operator on each atom where $\alpha$ stands for the direction index ($\alpha=x, y, z$). 

$\hat{\mathcal{H}}_\gamma$, an additional Rashba surface potential term, is induced by the deformation of the surface orbitals due to the local electric field at the interface. Indeed, due to the symmetry breaking at the surface, additional hopping interaction terms must be taken into account in the Hamiltonian at the two surface planes. We model such effect for the \textit{$sp^3$} basis by using the approach of Ast and Gierz~\cite{ast2012} for $\hat{\mathcal{H}}_\gamma$ considering three additional surface hopping terms: the so-called on-site $\Tilde{\gamma}_{sp_z}=\int d^3r \braket{is|\hat{z}|ip_z}$ energy, and respective $\gamma_{sp}=\int d^3r \braket{i~s|\hat{z}|(i+1)~p_z}$ and $\gamma_{pp}=\int d^3r \braket{i~p_x|\hat{z}|(i+1x)~p_z}$ hopping terms acting respectively between the $s-p_z$ and $p_x-p_z$ (or $p_y-p_z$) surface orbitals ($i$ is the atomic site at the surface, $i+1$ the lateral neighboring site and $1x$ a deplacement along the $x$ direction). These hopping terms are prohibited in the absence of surface electric field or electrical dipoles. We thus add the $\hat{\mathcal{H}}_\gamma$ Hamiltonian term of the form:

\begin{eqnarray}
\hat{\mathcal{H}_\gamma}=\left\{
\begin{array}{cc}
    \pm \gamma_{pp} \cos (\theta) & ~(i,j)\equiv (p_x,p_z) \\
    \pm \gamma_{pp} \sin (\theta) & ~(i,j)\equiv (p_y,p_z) \\
    \pm \gamma_{sp} &  (i,j)\equiv (s,p_z) \\
\end{array}
\right.
\end{eqnarray}
where the +(-) sign corresponds to the uppermost (lowermost) atomic plane and $\theta$ is the angle between the direction joining the two atoms considered and the $x$-direction. We then restrained ourselves to the in-plane surface hopping as for a pure 2-dimensional system. The best agreement with our ARPES results (Fig.~\ref{fig:2d-plots}a-b) is found for $\Tilde{\gamma}_{sp}=-0.2$~eV, $\gamma_{sp}$=0.3~eV and $\gamma_{pp}$=-0.6~eV for $x$=0.15 slightly departing from the values given for pure Bi, \textit{i.e.}~$\gamma_{sp1}$=0.45~eV and $\gamma_{pp}$=-0.27~eV~\cite{saito2016}, 
with opposite sign for the top and bottom surfaces due to the opposite direction of the potential gradient. We emphasize that these surface terms are required to reproduce at best the experimental electronic energy dispersion (Fig.~\ref{fig:2d-plots}a) involving TSS as well as the valence band subject to size quantification effects as well as subsequent Fermi surface shape (Fig.~\ref{fig:2d-plots}b) of a bare 12~BLs Bi$_{0.85}$Sb$_{0.15}$ film displaying both electron and hole pockets.

\vspace{0.1in}

The size of the Hamiltonian $\hat{\mathcal{H}}(k_x,k_y)$ to diagonalize is $16N \times 16N$ where $N$ is the number of bilayers (BLs). Once the Green function of the multilayer system is defined as:

\begin{equation}
    \hat{G}(\varepsilon,k_x,k_y)=\left[\varepsilon+i\delta -\hat{\mathcal{H}}(k_x,k_y)\right]^{-1}
\end{equation}

The partial density of state (DOS) $\mathcal{N}_\text{DOS}(\varepsilon)$ vs. the energy $\varepsilon$ equals $\mathcal{N}(n,\varepsilon)=-(1/\pi)~\text{Im} \text{Tr}[\hat{G}(\varepsilon,n,k_x,k_y)]$ whereas the spin density of states (spin-DOS) with spin along the $\alpha$ direction is $s_\alpha(\varepsilon)=(1/\pi)~\text{Im} \text{Tr}[\hat{\sigma}_\alpha \hat{G} (\varepsilon,k_x,k_y)]$. $\delta$ is the typical energy broadening ($\simeq$~50-100 meV) and the trace ($\text{Tr}$) is applied over the considered $sp^3$ orbitals on a given BL index ($n\in [1,N]$). The energy zero ($E=0$) refers to the Fermi level position. In the next, one will note $\mathbf{k}=(k_\parallel,k_z)$ the \textit{bulk} three-dimensional electronic wavevector when involved in the physical charge current production.

\vspace{0.1in}

The resulting calculation of the spin and orbital polarized surface band structure corresponding to the first BL of a 12 BLs thick Bi$_{0.85}$Sb$_{0.15}$ ($\simeq 5~$nm) are shown in Fig.~\ref{fig:2d-plots}c-e. Fig.~\ref{fig:2d-plots}c displays the spin-resolved DOS ($\sigma_y$) in the ($k_x$, $E$ energy) space below the Fermi energy ($E=0$) corresponding to the $\hat{y}$ in-plane spin component oriented along $\overline{\Gamma}-\overline{K}^\prime$ for a wavevector $k_x$ along the $\hat{x}=-\overline{M}|\overline{\Gamma}|\overline{M}$ of the 2-dimensional (2D) Brillouin zone (BZ). We observe respective positive (red) and negative (blue) spin projections for the two surface states $S_1$ and $S_2$ crossing the Fermi energy. This witnesses the spin momentum locking property with a maximum in-plane expected spin value $\bra{\psi_{\mathbf{k},n}}\hat{s}_y\ket{\psi_{\mathbf{k},n}}\simeq 0.3 \hbar=~0.6\left(\frac{\hbar}{2}\right)$ (spin polarization $\simeq$ 0.6) in average~\cite{Go2017}, close to $\overline{\Gamma}$. However, the resulting spin-resolved Fermi cut within the Brillouin zone (Fig.~\ref{fig:2d-plots}d) reveals the peculiarity and the complexity of the Fermi surface for Bi-based rhombohedral stacking constituted of: an almost circular central Rashba-like ring characterized by a spin-momentum locking with two additional holes and electrons pockets of hexagonal symmetry away from $\overline{\Gamma}$~\cite{Benia2015,Benia2015}. The ensemble of these TB calculations are compared with very good agreement to our (spin-) ARPES data (Figs~\ref{fig:2d-plots}a-b)~\cite{Baringthon2022,Rongione2023} portraying both the energy dispersion (a) and Fermi cut of the resolved density of states (b).  

\vspace{0.1in}

Equivalent analyses led for the $\hat{L}_y$ in-plane orbital angular momentum component is plotted on Fig.~\ref{fig:2d-plots}e-f in the same ($k_x,E$) window. One observes clear features corresponding to an orbital momentum locking with respective positive (red) and negative (blue) projections as for Bi$_2$Se$_3$~\cite{Zhang2013,Taniguchi2012}. Two main differences arise compared to the spin: \textit{i}) the orbit is polarized (locked) in the opposite direction in agreement with Ref.~\cite{Go2017} \textit{ii}) the orbital expectation value is of the same order of magnitude than the spin ($\bra{\psi_{\mathbf{k},n}}\hat{L}_y\ket{\psi_{\mathbf{k},n}}\simeq 0.4~\hbar$) in the range of values obtained for BiAg$_2$ as given by Ref.~\cite{Go2017}.  The resulting in plane orbital-resolved Fermi cut in the Brillouin zone is plotted on Fig.~\ref{fig:2d-plots}c emphasizing the relevance of the orbital polarization of the Fermi surface and more particularly of the electron pocket near $\overline{\Gamma}$. 

\section{Intrinsic Spin and Orbital Hall conductivities of BiSb alloys} 

We now turn to the evaluation of the intrinsic spin and orbital Hall conductivities of Bi$_{.85}$Sb$_{0.15}$ within the frame of the linear theory response. From \textcolor{blue}{SI-IIIA}, the respective \textit{intrinsic} spin (SHC) and orbital Hall (OHC) conductivities originating from the \textit{interband} coupling and scaling the bulk contribution writes:

\begin{equation}
    \sigma_{xy}^z=\frac{e^2\hbar}{\Omega}\sum_{\mathbf{k},n\neq l}\left(f_{\mathbf{k},n}-f_{\mathbf{k},l}\right)\frac{\braket{\psi_{\mathbf{k},n}|\hat{\mathcal{J}_y^z}|\psi_{\mathbf{k},l}}\braket{\psi_{\mathbf{k},l}|\hat{v}_x|\psi_{\mathbf{k},n}}}{(\varepsilon_{\mathbf{k}l}-\varepsilon_{\mathbf{k}n})^2} ,
    \label{SHC}
\end{equation}
where $\hat{\mathcal{J}}_y^z$ is either $\frac{1}{2}\{\hat{v}_y,\hat{S}_z\}$ or $\frac{1}{2}\{\hat{v}_y,\hat{L}_z\}$, the corresponding spin and orbital angular momentum current operator oriented along $\hat{z}$, $\hat{y}$ the flow direction of the angular momentum and $\hat{v}_x$ the velocity operator directed along the electric field $\mathcal{E}_x$. $\hat{\sigma}_{x,y,z}$ are evaluated for electronic states $\ket{\psi_{n,\mathbf{k}}}$ previously described, of eigenvalues $\varepsilon_n$ and $f_{\mathbf{k},n}$ is the Fermi-Dirac distribution, $e$ and $\Omega$ are the electronic charge and unit cell volume respectively. Any circular permutation between ($\hat{x},\hat{y},\hat{z}$) in Eq.~[\ref{SHC}] refers to a particular frame rotation.

\vspace{0.1in}

\begin{figure}[!t]
    \centering
    \includegraphics[width=\linewidth]{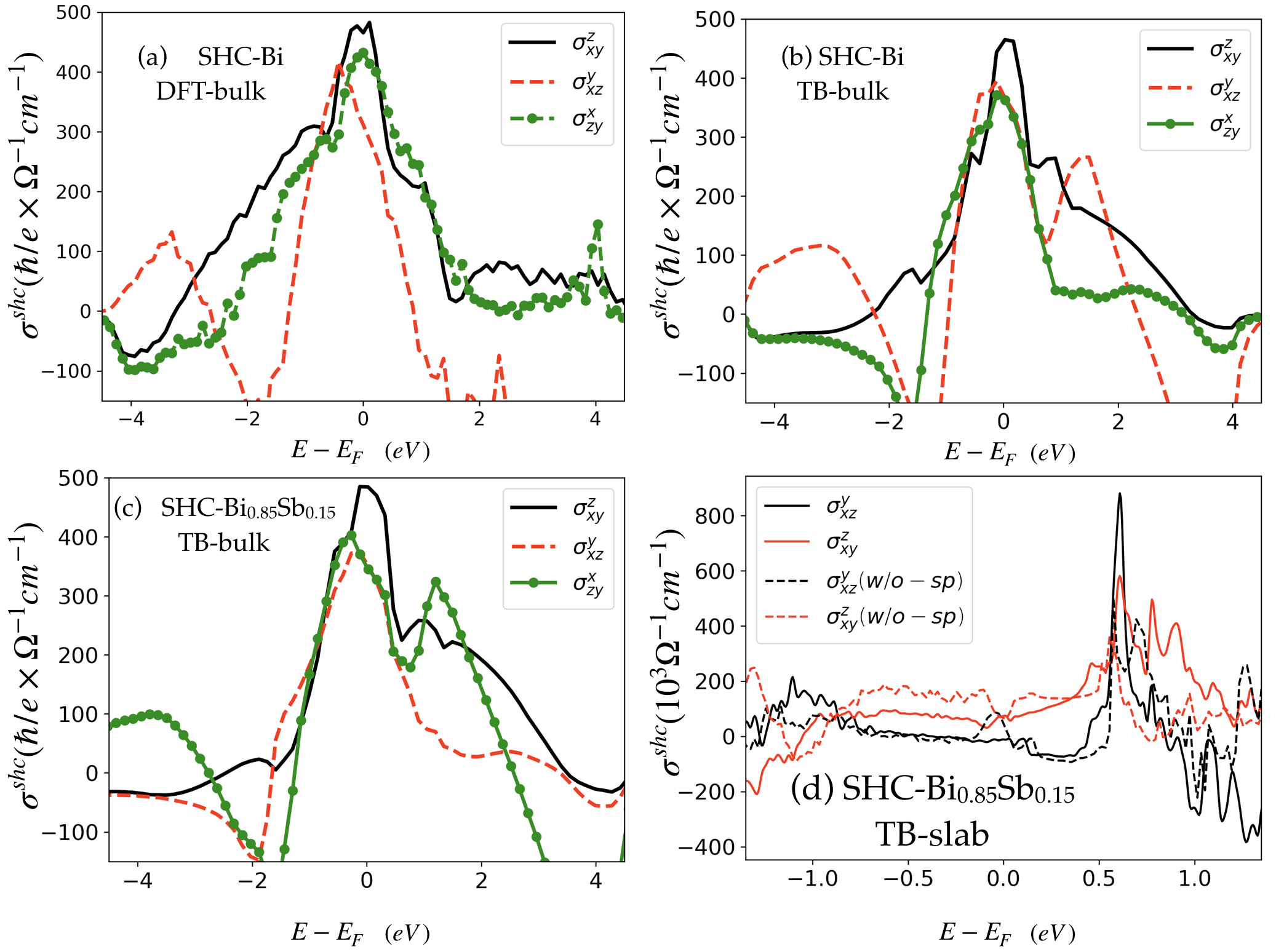}
    \caption{\textbf{SHC of Bi and Bi$_{0.85}$Sb$_{0.15}$ from DFT to TB}. SHC for bulk Bi (a,b). Three components of the SHC tensor vs. the position of the Fermi energy ($E$) obtained respectively by DFT (a) and by TB (b). SHC for Bi$_{0.85}$Sb$_{0.15}$ showing only moderate changes compared to pure Bi. In (d) we show the SHC components calculated for a TB 12 BLs (5~nm) Bi$_{0.85}$Sb$_{0.15}$ slab geometry for respective in-plane (red) and out-of-plane (black) spin currents $j^s$. We considered bare BiSb surface (dashed) and surface hopping term corrections (solid) with opposite values for the two opposite surfaces. In order to match the TB model with ARPES, we considered on-
    site \textit{s–p$_z$} coupling $\gamma_{sp}=-0.2$ eV, surface hopping terms respectively equalling $\gamma_{sp1}=0.3$~eV and $\gamma_{pp}=-0.6$~eV. The Fermi energy broadening was taken to 0.05~eV corresponding to a typical electronic relaxation time of about 10~fs.} 
    \label{fig:transport}
\end{figure}

Using Eq.~[\ref{SHC}] we have calculated both the SHC and OHC. Results for SHC are displayed in Figs.~\ref{fig:transport}, for pure Bi (Fig.~\ref{fig:transport}a-b) as well as Bi$_{0.85}Sb_{0.15}$ disordered alloy in the topological phase window (Fig.~\ref{fig:transport}c). The typical Fermi energy broadening has been chosen to be equal to 50~meV (typical electronic or spin scattering time $\tau_0$ onto the TSS of about 10 fs). 

\vspace{0.1in}

First, in order to validate our bulk TB parametrization, we compare the intrinsic response for pure Bi obtained from density functional theory (DFT) (Figs.~\ref{fig:transport}a) and from our TB bulk Hamiltonian (Figs.~\ref{fig:transport}b). For the DFT simulations we used SIESTA~\cite{siesta_method} as the code to calculate the self-consistent ground state. The out-of-equilibrium transport calculations use a full \textit{ab initio} DFT Hamiltonian matrices set obtained directly from the \textsc{SIESTA}~\cite{siesta_method}, employing atom-centered double-$\zeta$ plus polarization (DZP) basis sets.
We use the energy cutoff for real-space mesh of 400~Ry. The self-consistent spin-orbit potential is introduced \textit{via} the full off-site approximation~\cite{siesta_on-site_soc} using fully-relativistic norm-conserving pseudopotentials \cite{tm_pseudopotentials}. The system Hamiltonian and overlap matrices are obtained after performing a full self-consistent cycle and treated after withing a post-processing routine utilizing SISL~\cite{zerothi_sisl} as an interface tool.

\vspace{0.1in}

The intrinsic SHC obtained are observed to be in very close agreement with previous calculations~\cite{Sahin_BiSb} reaching however quite moderate values when compared to Pt, typically $\simeq 500~S/$cm within the chosen energy window close to the Fermi energy located at $E=0$ (Fig.~\ref{fig:transport}a-b). Indeed, three different components of the SHC tensor are displayed here revealing the rhombohedral crystal anisotropy. Fig.~\ref{fig:transport}c displays the corresponding spin-Hall conductivity for Bi$_{0.85}$Sb$_{0.15}$ showing only moderate changes compared to pure Bi. We have extended our approach to the multilayer slab geometry (\textcolor{blue}{see SI-IIB}). Our present calculations on a 12 bilayers (BL) Bi$_{0.85}$Sb$_{0.15}$ (Fig.~\ref{fig:transport}d) reveal that decreasing the BiSb layer thickness down to few BLs has a strong impact on the SHC in the bandgap region whatever the spin injection direction. Two types of calculations have been performed this way: with (w/ straight lines) and without (wo/ dashed lines) the additional surface hopping terms introduced by $\hat{\mathcal{H}}_\gamma$ as determined from our ARPES data as previously discussed. 

The severe decrease of the spin-Hall conductivity for ultrathin films down to less than $\simeq 100-200~S/$cm at the vicinity of the Fermi level ($E=0$) resulting from our calculations seems to originate from the changes in the electronic states at the surface and/or from quantization effects (owing to the $\left(\varepsilon_n-\varepsilon_n\right)$ increasing term in the denominator of Eq.~[\ref{fig:transport}]). Possibly, the top and bottom surfaces of opposite spin-chirality starts to mix together in the middle of the film leading to such severe SHC drop. We demonstrate then that SHC arising from evanescent TSS close to the gap can hardly drive an efficient spin-charge interconversion. SHC can be partially recovered in the CB for propagating states in the conduction band (CB) where localization effects are strongly attenuated.

\vspace{0.1in}

The extension of our TB theory to the orbital Hall conductivity (OHC) (\textcolor{blue}{see SI-IIIB}) shows that the latter ($\sigma_{OHE}^{BiSb}< 100~S/$cm) cannot explain neither the amount of the charge conversion at the level of Pt with $\sigma_{SHE}^{Pt}\simeq 2500-3000~S/$cm~\cite{jaffres2014} like observed in THz-TDS~\cite{Rongione2023,rho2023}. Such conclusions of non dominant SHC and OHC are even more supported by the additional reduction of the SCC expected at small layer thickness when the spin-diffusion length $\lambda_{sf}^{BiSb}$ is typically larger than some units of nm~\cite{Sharma2021} according to SCC~$\propto \frac{\left(\mathcal{G}_{\uparrow\downarrow}r_s\right)\tanh^2{\left(\frac{t^{BiSb}}{2\lambda_{sf}^{BiSb}}\right)}}{1+\left(\mathcal{G}_{\uparrow\downarrow}r_s\right)\coth{\left(\frac{t^{BiSb}}{\lambda_{sf}^{BiSb}}\right)}}\propto \left(\frac{t^{BiSb}}{2\lambda_{sf}^{BiSb}}\right)^2$~\cite{Rongione2023,Krishnia2024} with $t^{BiSb}$ the layer thickness, $r_s$ the spin resistance of BiSb and $\mathcal{G}_{\uparrow\downarrow}$ the spin-mixing conductance at Co$|$BiSb interfaces.

\section{Inverse Rashba-Edelstein (IREE) response} 

\subsection{Generalities about IREE}

Eventually, charge conversion processes occurring at interfaces may be better captured by Rashba-Edelstein effects (REE), where the inversion symmetry breaking may lead to angular momentum locking (spin or orbital). Either Hall effects or REE are usually attributed to different origins in terms of \textit{interband} vs. \textit{intraband} quantum transitions. Those are complementary in various phenomena as the spin-orbit torque (SOT) and its reciprocal effect as orbital pumping in both ferromagnetic resonance (FMR) or THz regimes. Inverse spin Rashba-Edelstein mechanisms (iSREE) have been already tackled in our previous work on Co$|$BiSb giving a reliable contribution to the ultrafast spin charge conversion ~\cite{Rongione2023,rho2023}. A constant charge conversion \textit{vs.} the layer thickness as observed was ascribed to the very short penetration of the TSS into the bulk in a sub-nanometric lengthscale. However, the missing ingredient here is the evaluation of the inverse orbital REE effect (iOREE). 

\vspace{0.1 in}

The calculation of the inverse Rashba-Edelstein tensor ($\Lambda^{iREE}$) corresponding to the reciprocal effect to the direct Rashba-Edelstein process (materialized by the $\kappa^{REE}$ tensor) is not straightforward in the linear response theory framework. The reason is that the external excitation is, now, not an extensive quantity (unlike the electric field $\mathcal{E}$ required for the calculation of the direct tensor) but an intensive quantity, presently the unbalanced angular momentum density (spin or orbital) carried by the excited (non-equilibrium) carriers. To that end, we generalize the expression of the REE length~\cite{Fert2013,Varotto2022,Johansson2023}, $\Lambda_{xy}^{iREE}$ obtained from a refined linear response theory~\cite{Rongione2023} and whose methodology is to evaluate at the same extend both charge current and angular momentum density response from \textit{time-varying magnetic-field $\mathbf{k}$-dependent pumping excitation}~\cite{shen2014} (\textcolor{blue}{SI-IVA}). The advantage of such approach is that the any anisotropy of electron scattering onto the Fermi surface related to related to both the spin and orbital momentum locking is automatically included there. One then obtains:

\begin{equation}
    \Lambda_{xy}^{iREE}=
\frac{\sum_{\mathbf{k},n}{\partial_\epsilon f_{\mathbf{k},n}}\bra{\psi_{\mathbf{k},n}}\hat{v}_{x}\tau_0\ket{\psi_{\mathbf{k},n}}\bra{\psi_{\mathbf{k},n}}\hat{\Pi}_y^{neq}\ket{\psi_{\mathbf{k},n}}}{\sum_{\mathbf{k},n}\partial_\epsilon f_{\mathbf{k},n}},
\label{IREE}
\end{equation}
with $\hat{\Pi}_y^{neq}$ the \textit{intraband} density matrix of the out-of equilibrium excited carriers impinging the surface and projecting onto the TI electronic states ($\psi_{\mathbf{k}, n}$) as previously introduced. The projection operation should be limited to a finite number of BLs only owing to the strong localization of the TSS, at least for Bi material~\cite{ishida2016}. We have introduced $\hat{y}$ the quantization axis along the magnetization/polarization of the ferromagnetic injector. $\Lambda_{xy}^{iREE}$ is weighted by the derivative of the Fermi distribution function $\partial_\epsilon f_{n,\mathbf{k}}=\partial_{E_{n,\mathbf{k}}} f_{n,\mathbf{k}}=\delta\left(E-\epsilon_{k,n}\right)$ that is the local density of states (DOS). Such expression for $\Lambda_{xy}^{iREE}$ is derived by considering a typical average relaxation $\tau_0$. Note that, discarding the orbital momentum and regarding spin-only states, one recovers the standard expression of the iSREE previously derived~\cite{Johansson2021,Johansson2023}. Eq.~[\ref{IREE}] is the general expression for $\Lambda^{iREE}$ parametrized by the projector $\Pi_y^{neq}$ characterizing the properties of the out-of equilibrium carriers the surface.

\subsection{Disantangling spin and orbital contribution to iREE}

We are now going to evaluate the action of the projector $\Pi_y^{neq}$ given in Eq.~[\ref{IREE}]. $\hat{\Pi}_y^{neq}$ appearing in Eq.~[\ref{IREE}] writes more generally as $\hat{\Pi}_y^{neq}=\Sigma_{m} \varrho_{m}\ket{\phi_{\mathbf{k},m}=\mu_\mathbf{k}\otimes s_\mathbf{k}}\bra{\phi_{\mathbf{k},m}=\mu_\mathbf{k}\otimes s_\mathbf{k}}$, where $\ket{\phi_{\mathbf{k},m}}$ contains both spin ($s_\mathbf{k}$) and orbital ($\mu_\mathbf{k}$) characters at any $\mathbf{k}$ point of the (non-equilibrium) entangled spin-orbital excited states. In the following, due to the various scattering and dephasing processes occurring between excitation and relaxation events within the nanometer thin layers, we will consider, from now on, the case of vanishing off-diagonal spin-density matrix elements impinging the surface/interfaces, \textit{i.e.} only statistics of pure spin and orbital states of eigenvectors parallel to the quantization axis. For the spin only case with polarization direction along $\hat{y}$, 
$\hat{\Pi}^{neq}_s=\frac{1}{2}\left(\varrho_{\uparrow}+\varrho_{\downarrow}\right)\hat{I}_{2\times 2}+\frac{1}{2}\left(\varrho_{\uparrow}-\varrho_{\downarrow}\right)\hat{\sigma_y}$, showing that the non-vanishing spin matrix element of $\bra{\psi_{\mathbf{k},n}}\hat{v}_{x}\tau_0\ket{\psi_{\mathbf{k},n}}\bra{\psi_{\mathbf{k},n}}\hat{\Pi}^{neq}_y\ket{\psi_{\mathbf{k},n}}$ in Eq.~[\ref{IREE}], symmetric in $\mathbf{k}$, only retains the $\left(\varrho_{\uparrow}-\varrho_{\downarrow}\right)\hat{\sigma}_y\propto\mathcal{P}_s \hat{\sigma}_y$ term as expected.

\vspace{0.1in}

According to those assumptions, in order to include both spin and OAM degrees of freedom, we will now consider \textit{separated}, however still possible \textit{correlated} physical quantities; and we are left with $\hat{\Pi}_y^{neq}=\hat{\Pi}^{neq}_s\otimes \hat{\Pi}^{neq}_{L}=\left(\varrho_{s_\mathbf{k}} \ket{s_\mathbf{k}}\bra{s_\mathbf{k}}\right)\otimes \left(\varrho_{\mu_\mathbf{k}}\ket{\mu_\mathbf{k}}\bra{\mu_\mathbf{k}}\right)$~\cite{Taniguchi2012}. $\ket{\mu_\mathbf{k}}\bra{\mu_\mathbf{k}}=\hat{\pi}_\mu$ are the orbital projectors onto respective $\mu=\pm 1,0$ \textit{p}-type orbital states~\cite{Go2017,Mokrousov2024} . 
Consequently, $\hat{\pi}_{\pm 1}=\ket{l_y=\pm 1}\bra{l_y=\pm 1}$ projects onto the orbital angular momentum $l_y=\pm 1$ along the in-plane $\hat{y}$ direction with respective $\pm 1$ eigenvalues whereas $\hat{\pi}_{0}=\ket{l_y=0}\bra{l_y=0}$ projects onto the \textit{p}$_y$ orbital ($l_y=0$) with:

\begin{eqnarray}
    \hat{\pi}_{\pm 1}=\frac{\pm\hat{L}_y}{2}(\hat{I}\pm \hat{L}_y) \quad ; \quad \hat{\pi}_{0}=\hat{I}-\hat{L}_y^2
    \label{projorbit}
\end{eqnarray}
which can be easily recovered from a generalized definition of projectors proposed in Ref.~\cite{graf2021}. One easily checks that $\hat{L}_y=\hat{\pi}_{+1}-\hat{\pi}_{-1}$, $\hat{L}_y^2=\hat{\pi}_{+1}+\hat{\pi}_{-1}$ and $\hat{\pi}_{-1}+\hat{\pi}_0+\hat{\pi}_1=\hat{I}$ the unity matrix for the determination of the iOREE in Eq.~[\ref{IREE}].    

\vspace{.1in}

The iREE charge current response to the incoming total angular momentum flow reads $\mathcal{J}_c\propto \Sigma_{\mathbf{k},n,}^{s,\mu} \left(\bra{\psi_{\mathbf{k},n}}\hat{v}_x\ket{\psi_{\mathbf{k},n}}\right)\times \left( \varrho_s\bra{\psi_{\mathbf{k},n}}\hat{\Pi}^{neq}_s\ket{\psi_{\mathbf{k},n}}\right)\times \left(\varrho_\mu \bra{\psi_{\mathbf{k},n}}\hat{\Pi}_L\ket{\psi_{\mathbf{k},n}}\right)$. To go beyond, we now have to introduce several definitions related to the spin-polarization $\mathcal{P}_s$, the orbital circular polarization $\mathcal{P}_c$ and the orbital linear polarization $\mathcal{P}_l$ with $\mathcal{P}_s=\left(\frac{\varrho_\uparrow-\varrho_\uparrow}{\varrho_\uparrow+\varrho_\uparrow}\right)$, $\mathcal{P}_c=\left(\frac{\varrho_{+1}-\varrho_{-1}}{\varrho_{+1}+\varrho_{-1}+\varrho_{0}}\right)$ and $\mathcal{P}_l=\frac{2\varrho_0-\varrho_{+1}-\varrho_{-1}}{2\left(\varrho_{+1}+\varrho_{-1}+\varrho_{0}\right)}$. We then have to inject those quantities into the matrix elements of the $\hat{\Pi}^{neq}_y=\left(\hat{\Pi}^{neq}_s\otimes\hat{\Pi}^{neq}_L\right)$ projection operator product in the expression of the IREE charge current response $\mathcal{J}_c\propto e\Sigma_{\mathbf{k},n,}^{s,\mu} \left(\bra{\psi_{\mathbf{k},n}}\hat{v}_x\ket{\psi_{\mathbf{k},n}}\right)\times \left( \bra{\psi_{\mathbf{k},n}}\left(\hat{\Pi}^{neq}_s\hat{\Pi}^{neq}_L\right)\ket{\psi_{\mathbf{k},n}}\right)$. 



\begin{figure*}[ht!]
\includegraphics[width=\linewidth]{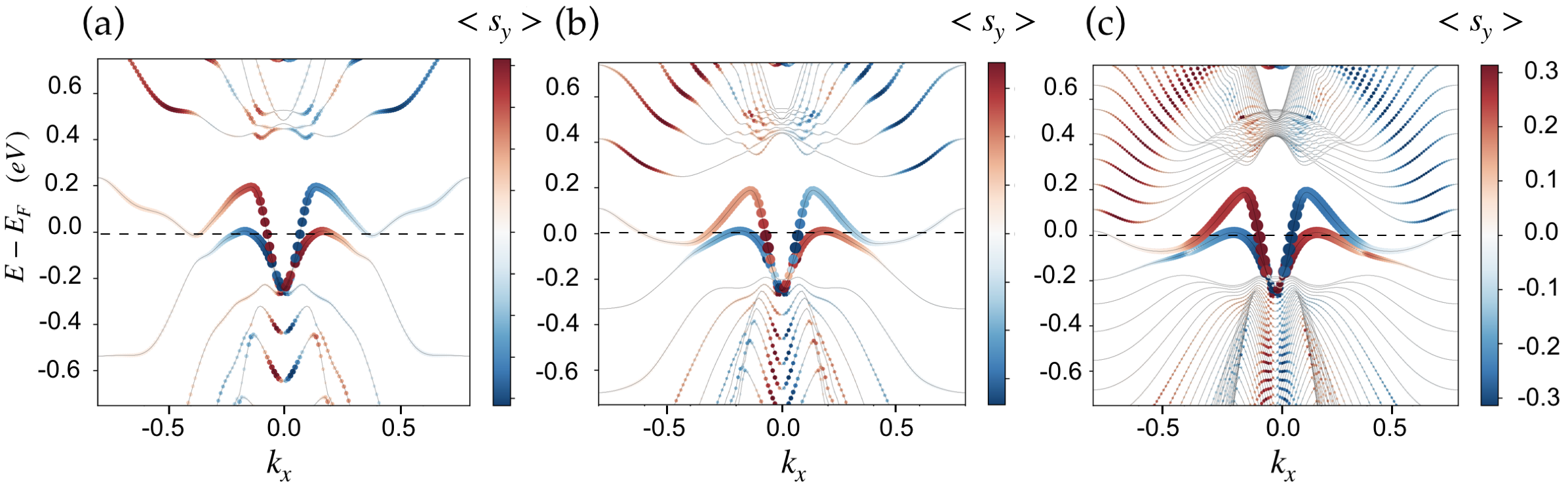}
\centering
\caption{\textbf{Electronic band structure of Bi$_{.85}$Sb$_{.15}$ at different thicknesses}. Electronic band structure of Bi$_{.85}$Sb$_{.15}$ calculated along the $\overline{\Gamma}-\overline{M}$ direction ($\hat{x}$) by TB methods corresponding respectively to a) 6 BLs (2.5~nm) b) 12 Bls (5~nm) and c) 36 BLs (15~nm). The color scale represent the projection onto the first BL of the spin-density along the $\hat{y}$ direction ($\braket{\hat{\sigma}_y}$) normal to $\overline{\Gamma}-\overline{M}$. The spin-polarization and spin momentum locking of the two surface state whose one is the TSS are clearly visible. One can clearly observe the quantification energy splitting in both the CB and VB as the film thickness is reduced. The surface Rashba hopping parameters used are given in the text. \textit{Values of $k_x$ in abscise are given in $\AA^{-1}$.}} \label{BiSb_thickness}
\end{figure*}

\begin{figure*}[ht!]
\includegraphics[width=\linewidth]{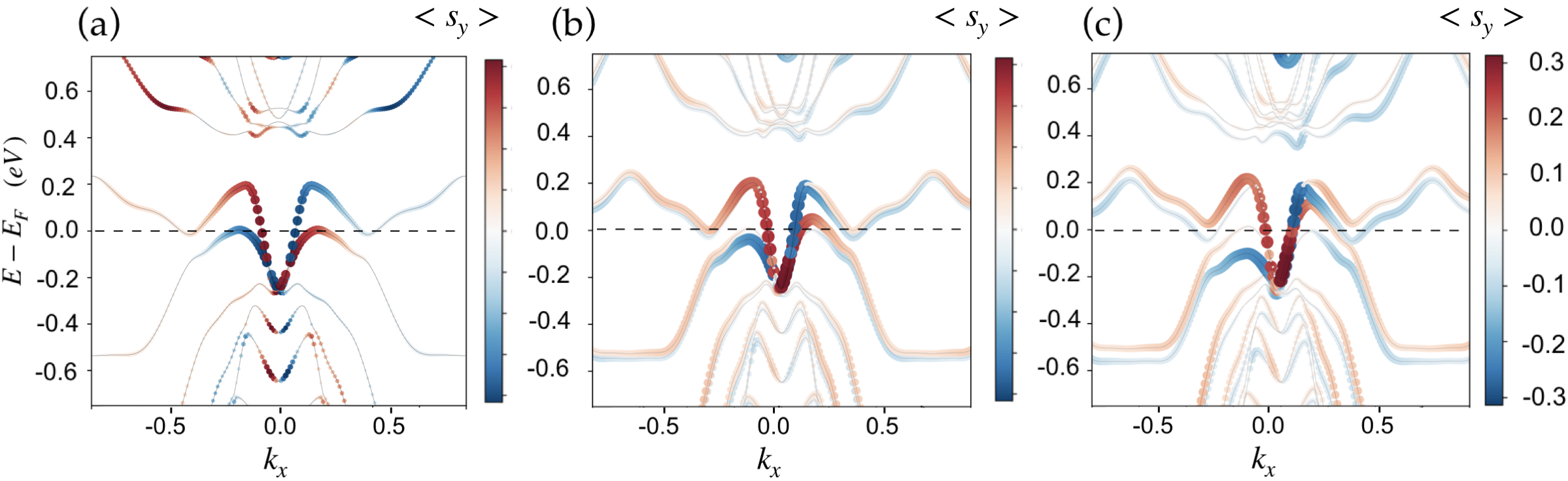}
\centering
\caption{\textbf{Electronic band structure of 12 BLs Bi$_{.85}$Sb$_{.15}$ with/without contact exchange}. Projection of the spin selected DOS of 12 BLs (5~nm) Bi$_{.85}$Sb$_{.15}$ calculated along the $\overline{\Gamma}-\overline{M}$ direction ($\hat{x}$) by TB methods corresponding respectively to zero interfacial exchange ($\Delta_{exc}=0~$eV), $\Delta_{exc}=0.2~$eV and $\Delta_{exc}=0.5~$eV applied in the sample plane at $45^o$ from the $\overline{\Gamma}-\overline{M}$ direction. The colored scale represent the projection onto the first BL of the electronic spin-density onto the $\hat{y}$ direction normal to $\overline{\Gamma}-\overline{M}$. The surface Rashba hopping parameters used are given in the text. The spin-polarization and SML of the two surface states, split by the exchange term and whose one is the TSS, are clearly visible. The lift of the degeneracy of the two bottom surface states is also clearly visible leading to attenuated spin-DOS on the first BL. \textit{Values of $k_x$ in abscise are given in $\AA^{-1}$.}} 
\label{BiSb_exchange}
\end{figure*}

\subsection{Specific properties of time-reversal symmetry}

The general expression of the $\hat{\Pi}_y^{neq}$ operator is given in the supplementary information. We are now going to discuss the more specific case of the time reversal symmetry (TRS) corresponding to the absence of any exchange interaction in the topological insulator. Because $\bra{\psi_{\mathbf{k},n}}\hat{v}_x\ket{\psi_{\mathbf{k},n}}$ is asymmetric in $\mathbf{k}$ upon TRS when $\mathbf{k}$ and $-\mathbf{k}$ are interchanged, there exists 3 different mechanisms to the charge conversion in the present case: \textit{i}) a pure spin-contribution giving rise to iSREE because $\braket{\hat{\sigma}_y}_{\mathbf{k},n}$ is also asymmetric in $\mathbf{k}$, \textit{ii}) a pure orbital contribution (iOREE) owing to that $\braket{\hat{\pi}_{-1}}_{\mathbf{k}}=-\braket{\hat{\pi}_{-1}}_{-\mathbf{k}}=-\braket{\hat{\pi}_{+1}}_{\mathbf{k}}$ is also asymmetric in $\mathbf{k}$ upon TRS, and \textit{iii}) an entangled spin-orbital contribution originating from the $\bra{\psi_{\mathbf{k},n}}\left(\hat{\pi}_0-\frac{\hat{L}_y^2}{2}\right)\hat{\sigma}_y\ket{\psi_{\mathbf{k},n}}=\bra{\psi_{\mathbf{k},n}}\left(\hat{I}_{3\times 3}-\frac{3}{2}\hat{L}_y^2\right)\hat{\sigma}_y\ket{\psi_{\mathbf{k},n}}$ term (\textcolor{blue}{see SI-IVC}). Note that $\braket{L_y^2}=\frac{2}{3}$ for no net orbital polarization. When TRS corresponding to non-magnetic systems (no exchange interaction) is applied to the general Eq.~\textcolor{blue}{[S13]} (\textcolor{blue}{SI-IVB}), we end up with only the antisymmetric part $\hat{\Pi}_y^A$ of $\hat{\Pi}_y$ reduced to three different terms:

\begin{widetext}
\begin{equation}
\begin{aligned}
\braket{\hat{\Pi}_y^A}_{\mathbf{k},n}= \frac{\varrho_T}{6}\left(
\mathcal{P}_s\braket{\hat{\sigma}_y\otimes \hat{I}_{3\times 3}}_{\mathbf{k},n} + \frac{3}{2}\mathcal{P}_c\braket{\hat{I}_{2\times 2}\otimes\hat{L}_y}_{\mathbf{k},n}+2\mathcal{P}_s\mathcal{P}_l\braket{\hat{\sigma}_y\otimes \left(\hat{I}_{3\times 3}-\frac{3}{2}\hat{L}_y^2\right)}_{\mathbf{k},n} \right)
\end{aligned}
\label{iree_trs}
\end{equation}
\end{widetext}
the spin (iSREE), orbital (iOREE) and spin-orbital iREE contributions parametrized by the respective polarization $\mathcal{P}_s, \mathcal{P}_c$  and $\mathcal{P}_l$ and that we are now going to evaluate in the following sections.

\vspace{.1in}

\section{Inverse spin REE contribution (iSREE): Case of unpolarized orbitals.}

Experimental data previously published~\cite{Rongione2023} clearly demonstrated an invariant THz emission and charge conversion \textit{vs.} Bi$_{1-x}$Sb$_x$ layer thickness and Sb content. This remarkable results discard the possibility of the spin-Hall conductivity as the main origin to the charge conversion as argued below. The latter process, calculated to be moderate in the present case, is expected to be very dependent on both the thickness and degree of disorder introduced by the Sb content. On the contrary, the iSREE is more dependent of the unique electronic state properties of the TSS which seems to be very robust \textit{vs.} layer thickness as shown in (Fig.~\ref{BiSb_thickness}a-c) as well as \textit{vs.}  the Sb content~\cite{Rongione2023}. Indeed, TSS display no remarkable difference in the expected spin values in the spin momentum locking structure and no large change in the related electronic dispersion (group velocity for instance). On the contrary, we can observe on Fig.~\ref{BiSb_thickness} that, although both the conduction (CB) and valence bands (VB) also acquire a certain spin polarization as the TSS do, those experience a large modification in their energy dispersion \textit{vs. BiSb thickness} with a subsequent change in their relative energy spacing originating from finite quantization effects. The typical energy splitting $\frac{\pi^2\hbar^2}{2m^* L_z^2}$ between two consecutive energy levels in the CB and VB equals about 0.04~eV for 36~BLs ($L_z=15$~nm) and 0.3~eV for 12~BLs ($L_z=5$~nm) in the range of what found previously~\cite{Baringthon2022} and also in agreement with Refs.~\cite{Ito2018,ito2020}. Nevertheless, the very large quantization energy experimented by 2.5~nm $Bi_{.85}Sb_{.15}$ larger than 0.5~eV seems to exclude the possibility of charge conversion on either CB and VB delocalized states while the transient THz signals display almost no difference~\cite{Rongione2023}.

\vspace{0.1in}

On the other hand, the localization depth of the TSS near the surface has been closely evaluated by DFT for pure Bi~\cite{ishida2016}. The localization depth in Bi hereafter seem to diverge only at the direct vicinity of the $\overline{M}$ point away from the zone center. Moreover, very recent literature demonstrated clear ARPES pictures of surface states acquired on 2~BLs and 4~BLs $Bi_{.85}Sb_{.15}$ revealing then the strong localization in depth of such TSS in BiSb alloy~\cite{palmstrom2024}. We have then checked that the ensemble of those properties remains valid for Bi$_{1-x}$Sb$_x$ in particular for $x=.15$ like studied here. Our present work clearly displays a strong localization of the two surfaces states for $Bi_{.85}Sb_{.15}$ near $\overline{\Gamma}$ over some units of atomic planes from a single atomic plane at $k_\parallel=0.1~k_{BZ}$ up to typically 4 atomic planes (localization over 2 BLs) at $k_\parallel=0.75 ~k_{BZ}$ with $k_{BZ}=\overline{\Gamma}-\overline{M}$ (\textcolor{blue}{see SI-IIB}) in strong agreement with Ref.~\cite{palmstrom2024}.

\subsection{iSREE in non-magnetic TI and TRS symmetry} 

We now turn to the evaluation of the inverse spin REE (iSREE) mainly related to the occurrence of the TSS in the gap. The pure spin iSREE arises from spin-polarized carriers, of polarisation $\mathcal{P}_s$ ($\varrho_\uparrow\neq \varrho_\downarrow$) and unpolarized orbital $\mathcal{P}_c=\mathcal{P}_l=0$, impinging the BiSb surface and generated by RF or laser spin pump. Therefore, $\hat{\Pi}_y^{neq}=\hat{\sigma}_y$ in Eq.~[\ref{IREE}] and the analytical form given in Refs.~\cite{Rongione2023, Johansson2023} is recovered.

\vspace{0.1in}

Our TB calculations of the iSREE length ($\Lambda_{xy}^{iSREE}$) for a scattering time $\tau_0=$10~fs are shown on Fig.~\ref{fig:orbital_response}a, for two different Bi$_{0.85}$Sb$_{0.15}$ thicknesses (respectively 6 and 12 BLs). The typical $\Lambda_{xy}^{SREE}\approx 0.2~$nm calculated near $E_F$ ($E=0$) approaches the $(\theta_{SHE}^{Pt}\times\lambda_{sf}^{Pt})$ product for Pt~\cite{jaffres2014} giving its SCC efficiency in the same length unit. The relative lack of variation of $\Lambda_{xy}^{iSREE}$ on the layer thickness~\cite{Rongione2023} can be explained by the short evanescent length of the surface states (TSS) in BiSb~\cite{Baringthon2022,Rongione2023,palmstrom2024} as for Bi~\cite{Ishida_2017}. This feature is exemplified in the inset of Fig.~\ref{fig:orbital_response}a showing the localisation of $\Lambda_{xy}^{iREE}$ at the surface over only 2~BLs within the 2D Brillouin zone near the $\Gamma$ point. Moreover, our previous work~\cite{Rongione2023} demonstrates no big differences to expect vs. the Sb content in Bi$_{1-x}$Sb$_x$ alloys keeping the full efficiency of spin to charge conversion in the topological window at the vicinity of the Fermi level.

\subsection{Case of proximity exchange with the 3\textit{d} FM: TRS breaking.} 

\begin{figure}
\includegraphics[width=\linewidth]{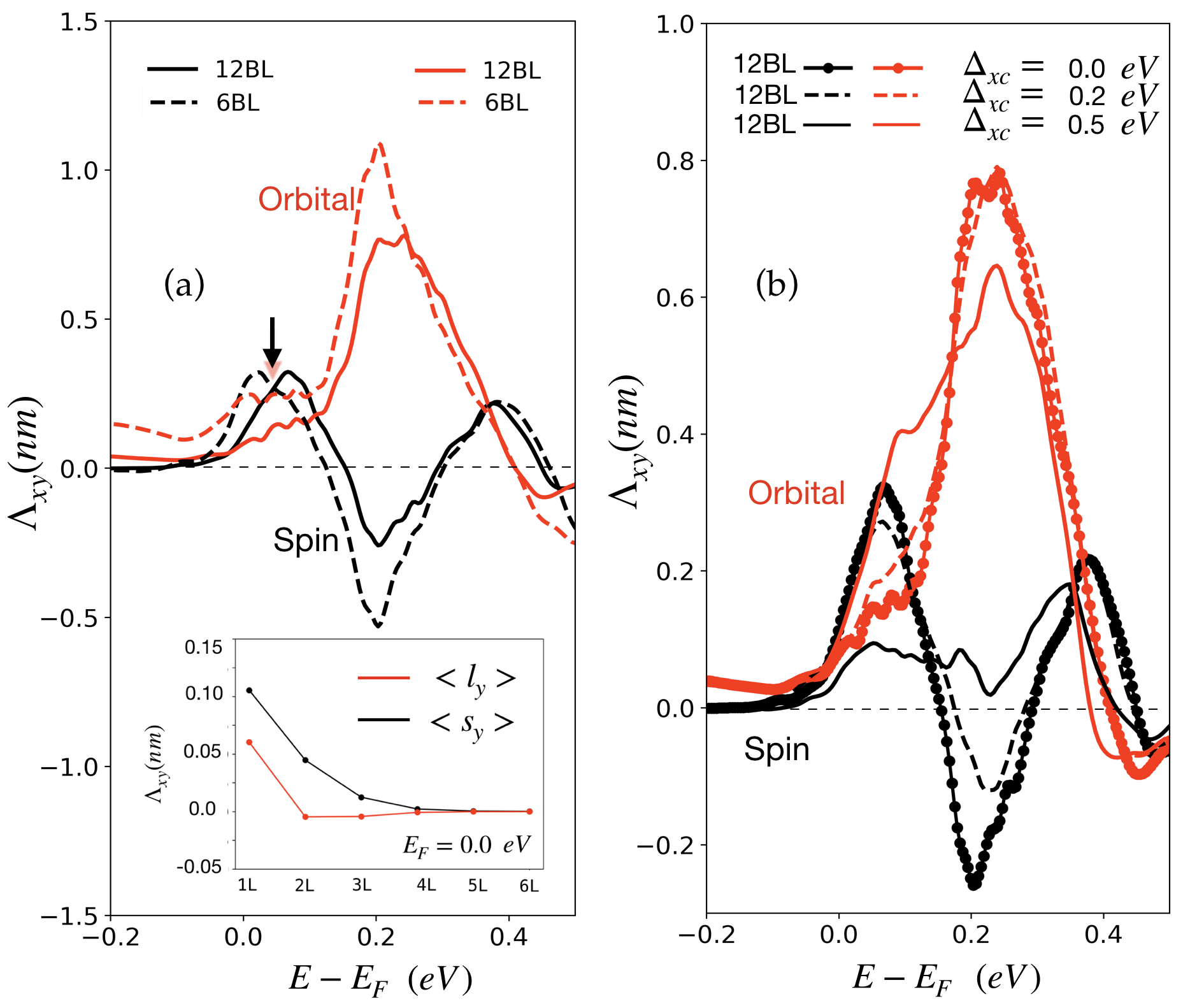}
\caption{\textbf{Linear response theory of the spin and orbital inverse REE.} iREE \textit{vs.} the Fermi energy position $E-E_F$ (a) for respective spin (iSREE) and orbital (iOREE) degrees of freedom displaying $\Lambda_{xy}^{iREE}$ (in $nm$) for 6 BLs (dashed lines) and 12 BLs (solid lines) Bi$_{.85}$Sb$_{.15}$. The charge interconversion is calculated for the top surface up by integration up to the middle of the layers. The inset shows the decay within each BL for the spin (black) and orbital (red) curves onto the TSS calculated at the Fermi level remarked by the vertical arrow. In (b) we show the IREE responses (spin and orbital) \textit{vs.} the Fermi energy position  in the case of an in-plane contact exchange of 0.2 (dashed) and 0.5~eV (dashed-point). 
}
\label{fig:orbital_response}
\end{figure}

We discuss now the case of a Co$|$BiSb interface possibly involving non-zero contact exchange in the first planes of BiSb at the interface with Co arising from \textit{sp-d} hybridization. We consider here a possible exchange process \textit{via} the occurrence of the surface Hamiltonian term $\hat{\mathcal{H}}=-\Delta_{exc}\hat{m}\cdot\hat{\sigma}$ parameterized by $\Delta_{exc}$ and acting on the first two atomic plane (first surface BL) of the BiSb film. By essence, the occurrence of such exchange terms has for effect to remove the TRS symmetry property without totally quenching the momentum locking properties (both spin and orbital) as the exchange strength $\Delta_{exc}$ remains typically smaller that the spin-orbit interactions. Fig.~\ref{BiSb_exchange}a-c displays the typical electronic dispersion and spin-DOS of a 5~nm thick (12 BLs) Bi$_{.85}$Sb$_{.15}$ layer respectively experimenting no exchange (a), a contact exchange strength $\Delta_{exc}=0.2$~eV (b) and  $\Delta_{exc}=0.5$~eV (c). The direction of the exchange (magnetization) has been set along the plane making an angle $\pi/4$ relative to the $\hat{x}$ and $\hat{y}$ directions. In details, the color code represents the expected value of the spin-related DOS projected along the $\hat{y}$ direction for electronic wavevector along $\hat{x}=-\overline{M}|\overline{\Gamma}|\overline{M}$ of the Brillouin zone and projected onto the first BL of BiSb (the one in contact with Co). In particular, one clearly observes the lift of the degeneracy of the two pairs of surface states of about 0.02~eV near $\overline{M}$ introduced by the exchange term, respectively the majority (blue) and minority (red) spin population, with however an evanescent spin-DOS character for the bottom ones (originating from opposite surface). Such a reduced value of 0.02~eV for the exchange energy splitting is explained by part by the localization of the exchange interaction onto the first BL and not distributed over the total BiSb thickness. As $\Delta_{exc}$ increases, Fig.~\ref{BiSb_exchange}a-c shows that \textit{i)} the energy of the spin-up states in red, opposite to the exchange direction, increases compared to the majoritary spin-down (blue) states. Moreover, \textit{ii)} the spin-momentum locking properties gradually disappears, in particular for the TSS occupying the higher energy bands without killing totally the one corresponding to the Rashba state S2 (energy dispersion dispersion in red color).
Upon such conditions, we have then evaluated the inverse spin Rashba-Edelstein response (iSREE) among other additional contributions to the spin-charge conversion due to the TRS breaking. The results (Fig.~\ref{fig:orbital_response}b) show that, although the iSREE is largely altered for $\Delta_{exc}=0.5$~eV, iSREE remains quite robust for moderate values of $\Delta_{exc}\simeq 0.2$~eV, leading thus to an important conclusion that moderate contact exchange at the surface of BiSb does not totally spoil the spin-conversion process.

\vspace{.1in}

\section{inverse orbital REE (ioREE)}  

We now turn on the discussion of the pure inverse orbital Rashba-Edelstein effect leading to orbital-to-charge conversion. According to Eq.~[\ref{iree_trs}], those processes may be parametrized by two different polarizations, either circular $\mathcal{P}_c$ or linear $\mathcal{P}_l$.

\subsection{iOREE: Case of polarized orbital involving circular OAM} 

One considers now the case of a circular orbital polarization $\varrho_{+1}\neq \varrho_{-1}$ giving rise to a non-zero $\mathcal{P}_c$ corresponding to an incoming OAM polarization (orbital accumulation) and $\mathcal{P}_s=0$ (no spin).
We then can use $\braket{\hat{\pi}_{+1}}_{\mathbf{k}}=-\braket{\hat{\pi}_{-1}}_{\mathbf{k}}$ and $\braket{\hat{\pi}_{+1}}_{\mathbf{k}}-\braket{\hat{\pi}_{-1}}_{\mathbf{k}}=\braket{\hat{L}_y}_{\mathbf{k}}$ in Eq.~[\ref{IREE}]. The orbital REE length, $\Lambda_{xy}^{iOREE}$, admits about the same value ($\Lambda_{xy}^{iOREE}\approx 0.15~$nm) than the spin in the bandgap at $E=0$ in Fig.~\ref{fig:orbital_response}a with the same typical evanescent length (inset of Fig.~\ref{fig:orbital_response}a), also at the level of the Pt spin-charge conversion efficiency. This gives rise to an additive contribution to the spin-charge conversion (same sign) at the Fermi level. This demonstrates the ability for an incoming pure orbital current to convert efficiently into a lateral surface charge current by iOREE.  
We also observe that the spin and orbital textures are opposite within the energy windows depicted in Fig.~\ref{fig:2d-plots}. 
This difference between spin and orbital behaviour is related to a different partition of the two S$_1$ and S$_2$ surface states \textit{iREE}. The comparison between spin and orbital responses are even more exemplified near the CB at higher energy ($E=0.2~$eV), where the orbital response is larger, of the order of $\lambda_{xy}^{iOREE}\simeq 0.8$~nm, than the spin one with an opposite sign. 

\vspace{0.1 in}

The orbital response appears to be even more robust than the spin over a large energy window (in the typical (0-0.3)~eV range as for hot carrier generated by a short pulse laser excitation~\cite{Rouzegar2022}), because of preserving the same sign of the charge conversion over the whole energy range. This is not the case for the spin component (ISREE) for which the sign of the spin-charge conversion seems to depend on the majority contribution of either S1 and S2 surface states. Such conclusion has nevertheless to be moderated: any linear-polarized laser optical excitation mainly generates spins, with subsequent OAM resulting from perturbation introduced by the spin-orbit coupling~\cite{Mokrousov2024} generally leading to a pretty small orbital polarization value $\mathcal{P}_c$. Also, the condition, not discussed here, is the absence of a strong decoherence between excitation processes and relaxation onto the TSS.

The general situation of correlated spin-orbital variables is then made possible \textit{via} the spin-orbit interactions acting in the 3\textit{d} transition metal (TM)ferromagnetic injector. It results a same qualitative expression for the iOREE as previously derived which amplitude is now scaled by the effective orbital polarization $\mathcal{P}_c=\zeta_c~\mathcal{P}_s $ as the product of the spin polarization $\mathcal{P}_s$ with $\zeta_c\propto \frac{\lambda_{SO}}{W}$ the orbital polarization appearing under the presence of the spin-orbit strength (scaled by $\lambda_{SO}$) ad where $W$ is the typical TM bandwidth (see \textcolor{blue}{SI-IVC}).

\vspace{0.1in}

To conclude this section, one can note that, when TRS is broken by an exchange field ($\Delta_{exc}$), the orbital response (iOREE) remains almost unaffected and maintains a large efficiency in the production of charge current (Fig.~\ref{fig:orbital_response}b). The main raison is that the exchange does not act directly on the OAM degree of freedom.

\subsection{Case of linear polarized orbital}. 


Now, we consider a non-zero orbital polarization free of contact exchange. For a certain linear polarization degree $\mathcal{P}_l\neq 0$ and non-zero spin polarization $\mathcal{P}_s$, an additional spin-orbital contribution originating from a linear orbital polarization of carriers is described by the $\hat{\Pi}_y^{neq}=\bra{\psi_{\mathbf{k},n}}\left(\hat{I}_{3\times 3}-\frac{3}{2}\hat{L}_y^2\right)\otimes \hat{\sigma}_y\ket{\psi_{\mathbf{k},n}}$ term which contribution on the TSS admits an opposite sign compared to the iSREE. For pure linear polarized orbital (only non-zero $\varrho_0$ component of the density matrix), the total spin-charge response is calculated to be pretty small in the bandgap considering the ($\Sigma_{\mathbf{k},n} \braket{\hat{\sigma_y}\hat{\pi}_0}_{\mathbf{k},n}$ term and larger in the CB region (\textcolor{blue}{see SI-IVC}). It originates from the orbital momentum locking and localization of the wavefunctions on the TSS projecting along the $l_y=\pm 1$ values forbidding thus the coupling to $l_y=0$ states. The propagating Bloch states are recovered in the CB partly releasing possible coupling with linear orbital $l_y=0$ states. 
The total \textit{spin-orbital} to charge conversion efficiency scales with the product $2\mathcal{P}_s \mathcal{P}_l$ as referring to the formula~[\ref{iree_trs}].

\section{Conclusions}

In summary, we have described the anatomy of respective spin and orbital-to-charge conversion in Bi$_{0.85}$Sb$_{0.15}$ TIs. By extending the analysis of the spin case to the orbital character, we were able to quantify both bulk spin and orbital contributions to the iS(O)HE, pointing out the small values of the total response. We have provided a detailed analysis of the Rashba-Edelstein effect arising from both degrees of angular momentum freedom. In the latter case, we disentangle the orbital contributions that lead to charge conversion such that, in terms of both circular and linear orbital polarization projectors. Such a decomposition allowed us to enhance a pure orbital part arising when such symmetry is broken by an in-plane magnetic exchange field. The use of orbital ferromagnetic injectors like played by Ni~\cite{Seifert2023,Xu2024} or CoPt~\cite{Xu2024b} alloys may be used to probe the orbital-charge conversion with BiSb alloys. One of the advantage of BiSb over others TIs like Bi$_2$Se$_3$ then resides in the strong localization of its TSS giving the possibility of charge conversion on very thin layer films embedded in spintronic devices emphasizing the strong interest for this material.

\vspace{0.1in}

\begin{acknowledgments}
The authors acknowledge P. Le Fèvre (Soleil Synchrotron, Saint-Aubin, France) and L. Baringthon for their help in the ARPES measurements. This study has been supported by the French National Research Agency under the project ’ORION’ ANR-20-CE30-0022-02, the project 'DYNTOP' ANR-22-CE30-0026 and by a France 2030 government grant managed by the French National Research Agency PEPR SPIN ANR-22-EXSP0009 (SPINTHEORY).
\end{acknowledgments}

\bibliography{bib1,bib2}

\end{document}